**Evolutionary Catastrophes and the Goldilocks Problem**

Milan M. Ćirković
*Astronomical Observatory of Belgrade*
*Volgina 7, 11160 Belgrade, Serbia*
*E-mail:* mcirkovic@aob.bg.ac.yu

**Abstract.** One of the mainstays of the controversial "rare Earth" hypothesis is the "Goldilocks problem" regarding various parameters describing a habitable planet, partially involving the role of mass extinctions and other catastrophic processes in biological evolution. Usually, this is construed as support for the uniqueness of the Earth's biosphere and intelligent human life. Here I argue that this is a misconstrual and that, on the contrary, observation-selection effects, when applied to catastrophic processes, make it very difficult for us to discern whether the terrestrial biosphere and evolutionary processes which created it are exceptional in the Milky Way or not. In particular, an anthropic overconfidence bias related to the temporal asymmetry of evolutionary processes appears when we try to straightforwardly estimate catastrophic risks from the past records on Earth. This agnosticism, in turn, supports the validity and significance of practical astrobiological and SETI research.

**Keywords**: mass extinctions – observation selection – evolutionary contingency – extraterrestrial life – catastrophism – anthropic principle

**1. Introduction**

One of the mainstays of the controversial "rare Earth" hypothesis in astrobiology is the "Goldilocks rule" regarding various parameters describing a habitable planet. If an arbitrary astrobiological parameter $\Omega$ can have various values on different habitable planets (let us neglect for the moment its temporal variation on a single planet), than only a limited range of values of $\Omega \in [\Omega_{min}, \Omega_{max}]$ will lead to a complex biosphere such as Earth's. In an instance of this rule, it is often suggested that the extent and rate of catastrophic events in the history of the terrestrial biosphere was both

- low enough to prevent the complete extinction of life or its developmental arrest at the microbial level, but
- high enough to provide sufficient "evolution pump" and enable the faunal overturns and opening ecological niches for more complex and advanced evolutionary designs.

The reasoning here is that, on an average, the terrestrial-planet rate of catastrophes is going to be either higher or lower than the rate characterizing Earth's history. Therefore, the conditions necessary for the emergence of complex life and intelligent observers are bound to be quite rare. This has been argued by Ward and Brownlee (2000) in the "Bible" of the "rare Earth" movement, and is essentially accepted by all supporters of this astrobiological view (particularly illustrative for different reasons are Carter 1993; Webb 2002; Conway Morris 2003; and various writings of Frank Tipler, e.g., Tipler 2003). Catastrophes we are dealing with here include not only well-publicized ones like the asteroidal/cometary impacts and supervolcanic eruptions, but also those less-understood like a close supernova/γ-ray burst or even those nobody has suggested thus far (Leslie 1996; Bostrom 2002).

In comparison to some other claims of the "rare Earth" supporters, this one sounds plausible and even appealing. After all, only in the last quarter of century we have realized the significance of the role mass extinctions play in determining courses of evolution, especially after the study of Alvarez et al. (1980); among fine popular reviews are Raup (1991), Courtillot (1999), and Erwin (2006). Symbolically, the reasoning involved can be presented as

(1) Earth possesses property X
(2) X seems *a priori* unlikely among Earth-like planets
(3) Our emergence is contingent on X
(4) We are intelligent observers
-------------------------------------------------------------------------------------------------
The emergence of intelligent observers on other Earth-like planets is unlikely.

X in this particular case can be defined as "fine-tuned catastrophism". This reasoning can be criticized from several different viewpoints. For instance, one could deny premise (3), an approach which is unpopular today due to theistic misuses in the past, although it is denied by some secular philosophers as well (e.g. Wright 2000). Often, people question the *ceteris paribus* clause in (2). I shall take a different approach: first, I wish to argue that (3) should be rephrased by emphasizing its **temporal** nature:

(3)' Our emergence **at present epoch** is contingent on X

This has multifold advantages, the most important being the possibility of connecting with the well-studied evolutionary processes, both on Earth and in the Milky Way. In addition, it serves to underscore that the argument pertains essentially to us, *Homo sapiens*, no matter how its authors pretend otherwise (in order to escape the charge of anthropocentrism; e.g. Carter 1983).
    Next question we should ask is: how big a catastrophe at each point in the history of Earth is consistent with the existence of intelligent observers at $t = t_0$? Notice that the observers we are asking here about need not be **us,** humans, in either morphological, phylogenetical or just chronological way. Without entering the hard and heated topic to what extent is our presence today contingent or convergent (for different views see Gould 1989; McShea 1998; Conway Morris 1998, 2003; Shanahan 1999, 2001), we can attempt to draw some preliminary conclusions by investigating first a specific catastrophic event, and subsequently showing that such very destructive episodes in fact undermine the information value of past records. This, I shall try to show, makes the Goldilocks argument misleading, since it does not take into account important observation selection effects.

**2. Catastrophes and Observation Selection**

Let us consider a case of the catastrophic event which has influenced evolution of Earth's biosphere beyond serious doubt: the Chixhulub impact of ~65 Myr ago. Since Alvarez et al. (1980) proposed an impact as the major causal agency of the well-known mass extinction episode at the Cretaceous/Tertiary boundary, the scientific case for impact has been strengthened almost continuously. Nowadays, we know that the size of the impactor was between 10 and 20 km, depending on parameters such as the internal composition, the incident angle and the velocity of the impacting body (Hughes 2003). But impacting bodies can, in principle, be of widely varying sizes; after all, micrometeorites bombard Earth all the time, and larger particles create beautiful meteor showers apparently without threatening the biosphere or in any known way influencing the evolutionary processes. On the other hand, studies of early history of the Solar System suggested that collisions with bodies hundreds of kilometers in size remaining at that epoch had caused repeated meltdown of the entire planetary crust and perhaps even complete atmosphere blow-off (Maher and Stevenson 1988). Thus, only a finite—and quite small—range of impactors at the fixed epoch of K-T boundary could have caused the evolution of modern humans. A schematic presentation of this effect is shown in Figure 1, which considers the range of possible impactors at the K-T boundary. Only impactors in the roughly shadowed region of size distribution could lead to the emergence of humans. (Whether some other form of intelligence could emerge without the mass extinction for general biological reasons is highly uncertain; on the other hand, it seems quite unlikely that this would have occurred at about the same **time**.)

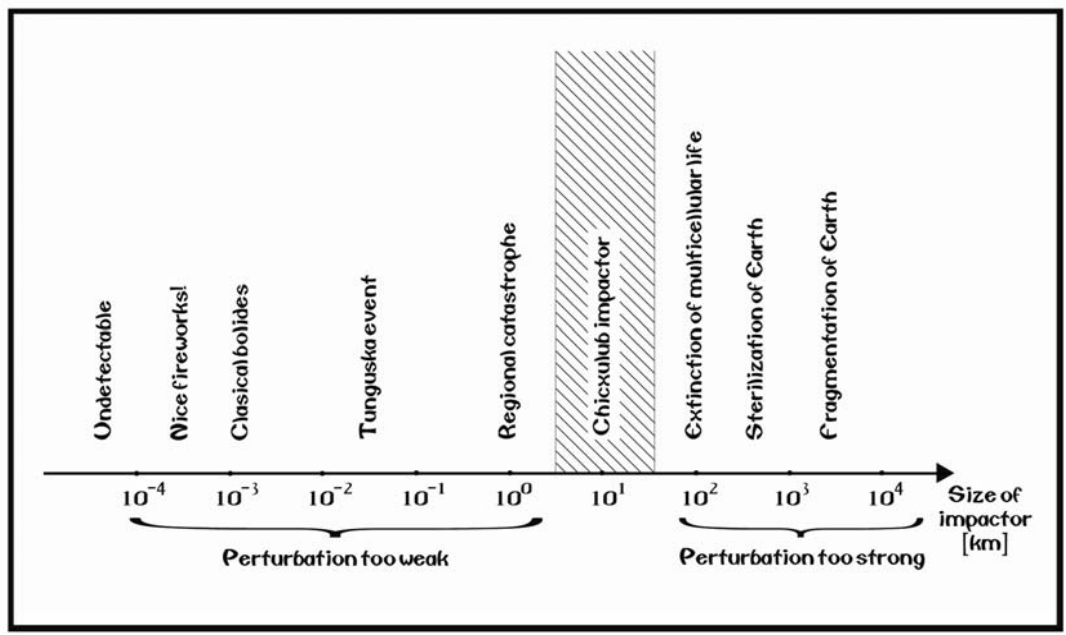

**Figure 1:** Goldilocks problem at the K/T boundary: only impactors in the narrow shaded strip would have led to the "right" mass extinction (leading to the development of humans at present epoch). **Both** smaller and larger impacts would have resulted in non-existence of humans. This remains valid even if we acknowledge the existence of the empirical upper limit on the size of an impactor.

Obviously, the same "Goldilocks problem" reasoning applies to other physical causative agents of catastrophic events. But, what does this mean? Do we have to conclude that only an overwhelming, *a priori* extremely improbable instance of luck brought about catastrophes fine-tuned for our existence, as the "rare Earth" proponents seemingly suggest? Not really, since (i) it is obviously incorrect to conclude that in the absence of humans **no** intelligent observers would have arisen by this date (cf. Russell 1983; McKay 1996); and (ii) at least those catastrophes which are stochastic in the epistemic sense are subject to a simple observation-selection effect which I shall now illustrate by a toy model.

Consider the simplest case of a single very destructive global catastrophe, for instance, a Toba-like supervolcanic eruption.[1] The evidence we take into account in a Bayesian manner is the fact of our existence at the present epoch (not necessarily including the influence of the perturbation on the evolutionary pathways; this is just a binary toy model). We can schematically represent the situation in Figure 2: *a priori* probability of catastrophe is *P* and the probability of human survival (or an insufficiently strong perturbation leaving evolutionary pathways within the morphological subspace containing humans) of the catastrophic event is *Q*. We shall suppose that the two probabilities are (i) constant, (ii) adequately normalized and (iii) applicable to a particular well-defined interval of past time. Event $B_2$ is the occurrence of the catastrophe, and by *E* we denote the evidence of our present existence.

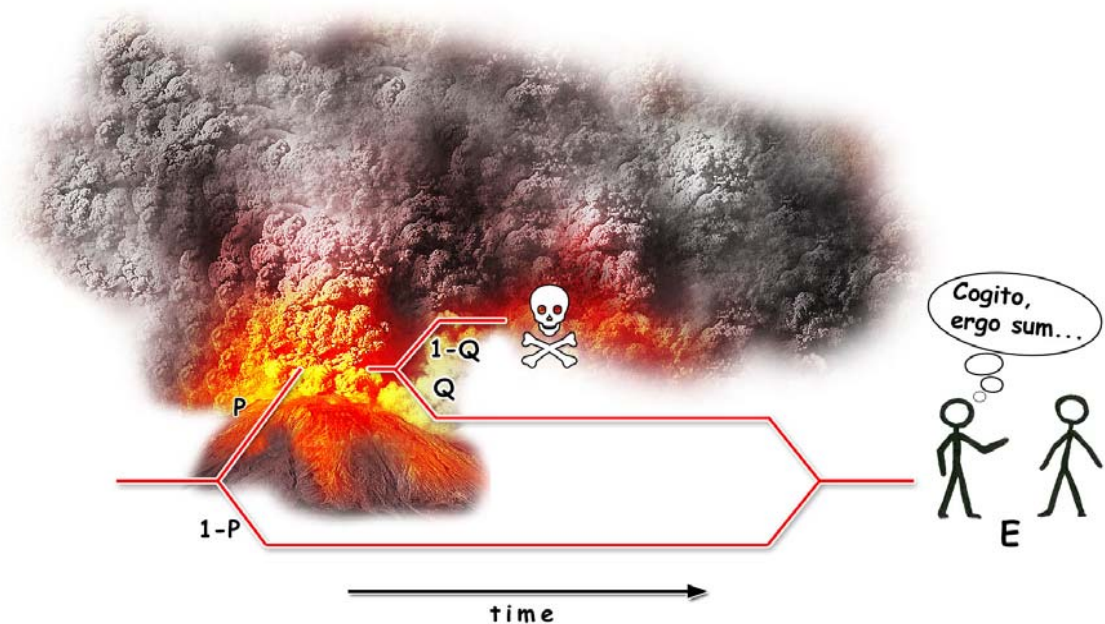

**Figure 2:** A schematical presentation of the single-event toy model. The evidence E consists in our present-day existence.

The direct application of the Bayes formula in form

$$P(B_2 \mid E) = \frac{P(B_2)P(E \mid B_2)}{P(B_1)P(E \mid B_1) + P(B_2)P(E \mid B_2)},\qquad (2)$$

using our notation, obtains the *a posteriori* probability as

$$P(B_2 | E) = \frac{PQ}{(1-P)\cdot 1 + PQ} = \frac{PQ}{1-P+PQ}.$$ (3)

---

[1] Supervolcanism gives us an example of an almost-too-strong environmental stress: Toba supereruption (Sumatra, Indonesia, 74 Kyr BP) may have, according to at least one speculative hypothesis, reduced human population to ~1000 individuals, nearly causing the extinction of humanity (Rampino and Self 1992; Ambrose 1998). Even if that is shown to be wrong by subsequent research, it seems clear that the high correlation of supervolcanic eruptions with the mass extinction episodes testifies on their evoluionary importance.

Only a rather straightforward algebraic manipulation shows that

$$P(B_2 | E) \leq P,$$ (4)

i.e., that we tend to underestimate the true catastrophic risk. It is intuitively clear why: the symmetry between past and future is broken by the existence of an evolutionary process leading to our emergence as observers at this particular epoch in time. We can expect a large catastrophe tomorrow, but we cannot—even without any empirical knowledge—expect to find traces of a large catastrophe which occured yesterday, since it would have preempted our existence today.

We can define the *anthropic overconfidence parameter* as

$$\eta \equiv \frac{P(\text{a priori})}{P(\text{a posteriori})},$$ (5)

illuminating the magnitude of this bias in a quantitative way; obviously, our inferences from the past become unreliable if $\eta > 1$ and rather useless for $\eta \gg 1$. In the special case of our toy model, the overconfidence parameter becomes

$$\eta = \frac{P}{P(B_2 | E)} = \frac{1-P+PQ}{Q}.$$ (6)

All values larger than $\eta \sim 1$ indicate that we are seriously underestimating the extinction risk. However, even quite conservative numerical values give rather depressing results here. If, for instance, we take $Q = 0.1$, $P = 0.5$ (corresponding to a fair-coin-toss chance that an event similar or slightly larger than the Toba supereruption occurs once per ~ 1 Myr of human evolution), the resulting value of the overconfidence parameter is $\eta = 5.5$, indicating a gross error in the risk estimates! Values of overconfidence as a function of severity (as measured by the extinction probability $1 - Q$) are shown in Figure 3.

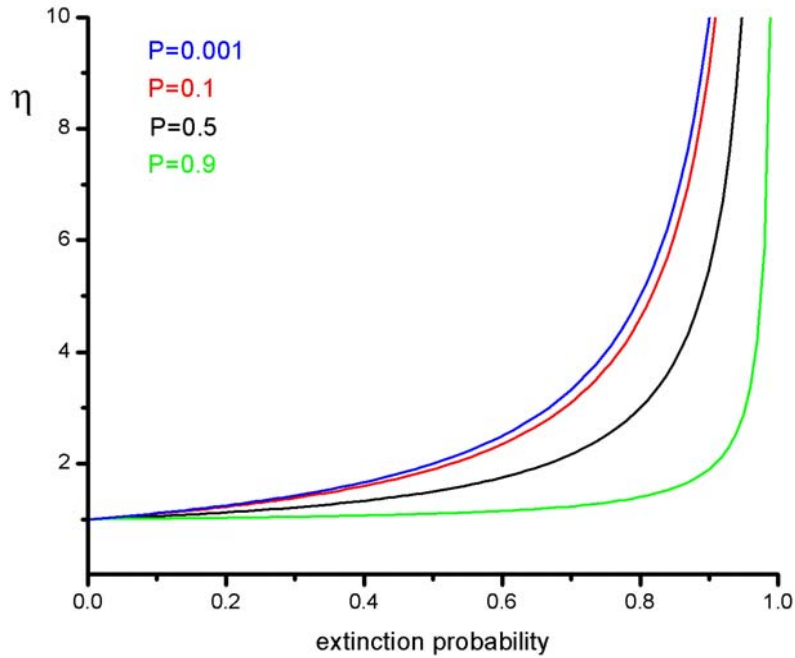

**Figure 3:** Overconfidence parameter as a function of the extinction probability 1-Q in our single-event toy model. Different values of the real event probability density *P* (appropriately averaged) are color-coded. We notice that the overconfidence bias is strongest for low-probability density ("rare") events.

Note that

$$\lim_{Q \to 0} \eta = \infty . \tag{6}$$

*Overconfidence becomes very large for very destructive events!* An obvious consequence is that the possibility of absolutely destructive events, which humanity has no chance of surviving at all ($Q = 0$) completely destroys our confidence in predicting from past occurrences. This almost trivial conclusion is not, however, widely appreciated. On the contrary, a rather well-known argument of Hut and Rees (1983) on the vacuum phase-transition contains the not-so-trivial error of not taking observation selection into account. Of course, a more sophisticated model involving series of random catastrophic events with various causes need to be developed, but the main philosophical point is clear: we cannot reason as if our past evolution is truly typical for a terrestrial planet without taking into account our present existence.

This has important consequences for our study of the role of catastrophic events in the general astrobiological context, i.e. in the evolution of an average biosphere in the Galaxy. Is evolution on the other habitable planets in the Milky Way influenced more or less by catastrophes? We cannot say, because the stronger catastrophic stress is (the larger analog of our probability $1 - Q$ is on the average), the less useful information can we extract about the proximity—or else—of our particular historical experience to what is generally to be expected. Only serious astrobiological studies can give a useful answer to this conundrum. Some data already exist. For instance, one well-studied case is the system of the famous nearby Sun-like star Tau Ceti which contains both planets and a massive debris disk, analogous to the Solar System Kuiper belt. Modeling of Tau Ceti's dust disk observations indicate, however, that the mass of the colliding bodies up to 10 kilometers in size may total around 1.2 Earth-masses, compared with 0.1 Earth-masses estimated to be in the Solar System's Edgeworth-Kuiper Belt (Greaves et al. 2004). Thus, Tau Ceti's dust disk may have around 10 times more cometary and asteroidal material than is currently found in the Solar System – in spite of the fact that **Tau Ceti seems to be about twice as old as the Sun**. Why the Tau Ceti System would have a more massive cometary disk than the Solar System is not fully understood, but it is reasonable to conjecture that any hypothetical terrestrial planet of this extrasolar planetary system is subjected to much more severe impact stress than Earth has been during the course of its geological and biological history.[2]

## 3. Discussion

I have considered the influence of observation-selection on our thinking about role of catastrophes in the evolutionary history of Earth and, by analogy, other habitable terrestrial planets in the Galaxy. Two points are of particular importance for future discussions:

(1) We cannot hope to realistically assess the **general** importance of catastrophic events in shaping the evolution on habitable planets as long as we are limited to the **past local** records.
(2) Invoking fine-tuning of catastrophes to support the "rare Earth" hypothesis is misguided, since any information one infers from the past records is either erased or heavily skewed by the observation-selection effects.

It is significant to note that, in principle, it is quite possible that we need to revise our views on the general frequency of catastrophes upward, and thus the universe might be more hostile to life throughout and the probability of **escaping** catastrophic extinction is very small.[3] This is still consistent with the conclusion (1) above. This possibility should be always kept in mind, at least as a cautioning note for the overwhelming enthusiasm similar to that reigning about SETI in 1960s and 1970s. However, there is a host of different reasons for taking a more moderate position, somewhere between the extremes of naive contact-optimism and anthropocentric skepticism of "rare-Earthers" (Dyson 1966; Gould 1987; Dick 2003; Ćirković and Bradbury 2006).

There are several practical consequences of conclusions (1) and (2). One is certainly that only astrobiology, currently undergoing explosive development, will be able to tell us whether there are other biospheres and other intelligent communities in the Milky Way. No amount of armchair theorizing can escape the observation selection effects related to the evolutionary development of intelligent observers on Earth. In contrast, fruitful studies like those of Greaves et al. on Tau Ceti and other exoplanetary systems, offer us a glimpse of the "true" underlying physical background of the astrobiological evolution in the Milky Way. Besides, important philosophical questions related to the foundations of astrobiology need to be addressed. In particular, the observation-selection effects, often demeaned under the heading of anthropic principle(s), need to be reevaluated, in particular in views of its potentially important impact on all studies of future catastrophic risk facing an Earth-confined humanity (Bostrom and Ćirković 2008).

---

[2] Earth-like planets have not been discovered yet around Tau Ceti, but in view of the crude observational techniques employed so far, it has not been expected; the new generation of planet-searching instruments currently in preparation (DARWIN, Gaia, TPF, etc.) will hopefully settle this problem.

[3] On the other hand, one should not underrate the constructive aspect of global catastrophes: oft-mentioned "evolution pump" enabling faunal overturns and opening hitherto inaccessible parts of the evolutionary morphospace. This effect is, of course, extraordinarily hard to quantify (e.g. Kitchell and Pena 1984; Benton 1995). The reasoning similar to the one above applies here too: any observer is bound to find—upon her developing evolutionary biology and paleontology in sufficient detail—that the evolutionary pump was sufficiently strong in her past. This is valid even if *a priori* rate of catastrophic events is on the average significantly **lower** than that commonly inferred from Earth's past record.

**Acknowledgements.** I use the opportunity to thank the Future of Humanity Institute, Oxford, UK, for the kind hospitality during the period this paper was conceived. This work has been partially supported by the Ministry of Science of the Republic of Serbia through the project ON146012. Useful discussions with Anders Sandberg, Nick Bostrom, Robert Bradbury, Slobodan Popović, and Robin Hanson are also hereby acknowledged. Two referees for the *International Journal of Astrobiology* have provided helpful comments on a previous version of the manuscript. Aleksandar Zorkić is credited for help in drawing Figure 2. Invaluable technical help and kind encouragement of Irena Diklić have been instrumental in the completion of this project.